# What Drives Virtual Influencer's Impact?

Giovanni Luca Cascio Rizzo

Jonah Berger

Francisco Villarroel Ordenes

Giovanni Luca Cascio Rizzo (glcasciorizzo@luiss.it) is a doctoral student at the LUISS Guido Carli University. Jonah Berger (jberger@wharton.upenn.edu) is an associate professor of marketing at the Wharton School at the University of Pennsylvania. Francisco Villarroel Ordenes (fvillarroel@luiss.it) is an assistant professor of marketing at the LUISS Guido Carli University. The authors thank Stefano Puntoni for helpful feedback on the manuscript.

# ABSTRACT

In the midst of the influencer marketing boom, more and more companies are shifting resources from real to virtual (or computer-generated) influencers. But while virtual influencers have the potential to engage consumers and drive action, some posts resonate and boost sales, while others do not. What makes some virtual influencer posts more impactful? This work examines how including someone else in photos shapes consumer responses to virtual influencers' posts. A multimethod investigation, combining automated image and text analysis of thousands of social media posts with controlled experiments, demonstrates that companion presence boosts impact. These effects are driven by trust. Companion presence makes virtual influencers seem more human, which makes them seem more trustworthy, and thus increases the impact of their posts. Taken together, the findings shed light on how others' presence shapes responses to virtual influencer content, reveal a psychological mechanism through which companions affect consumer perceptions, and provide actionable insights for designing more impactful social media content.

Influencers have become a key part of marketing strategy. Rather than watching television, or consuming offline media, consumers are spending more and more time online, particularly on social media (Lee and Junqué De Fortuny 2021). Further, consumers are wary of ads, so turn to others' opinions as a way to collect information and make decisions (Wies, Bleier, and Edeling 2022). Consequently, more than 75% of companies are now investing in online influencers as a way to raise awareness, encourage consideration, and drive purchase (Leung et al. 2022a).

In the midst of this boom, though, a new type of influencer has emerged. Given how carefully influencers craft online identities, creators have realized that influencers don't actually have to be real people at all. As a result, marketers are increasingly using virtual, or computer-generated influencers, to achieve their goals (Appel et al. 2020; Leung, Gu, and Palmatier 2022a). These influencers post images and content like people, but creators can control how they look, dress, and act. Over 25 million consumers follow virtual influencer Lu do Magalu, for example, viewing pictures of her wearing Adidas or supporting certain soccer teams. Similarly, virtual influencer Lil Miquela's 2020 Samsung #TeamGalaxy campaign generated over 126 million organic views and increased brand mentions by 12% (P2P 2022). This approach is growing rapidly, and the market is expected to reach $22 billion by 2025 (WGSN 2022).

But while some virtual influencers' posts garner lots of engagement and drive action, others do not. Why? And are there things virtual influencers can do to increase their impact?

We suggest that the presence of others plays an important role. Rather than sharing an image[1] of them eating alone, for example, a virtual influencer endorsing a restaurant could share a picture of them eating with friends (see figure 1). We suggest that showing them with others can make virtual influencers seem more like real people, which should make them seem more trustworthy, and thus increase engagement with, and impact of, their posts.

---
[1] We use "image", "picture", and "photo" as interchangeably throughout.

A multimethod investigation, combining text and image analysis of almost 10,000 virtual influencer posts with controlled experiments, test these possibilities. They demonstrate that including someone else in a photo (i.e., what we describe as companion presence) increases engagement and likelihood of choosing the endorsed product. Further, they highlight the underlying role of trust (and anthropomorphism) in driving these effects.

**FIGURE 1:** EXAMPLE IMAGES OF COMPANIONS

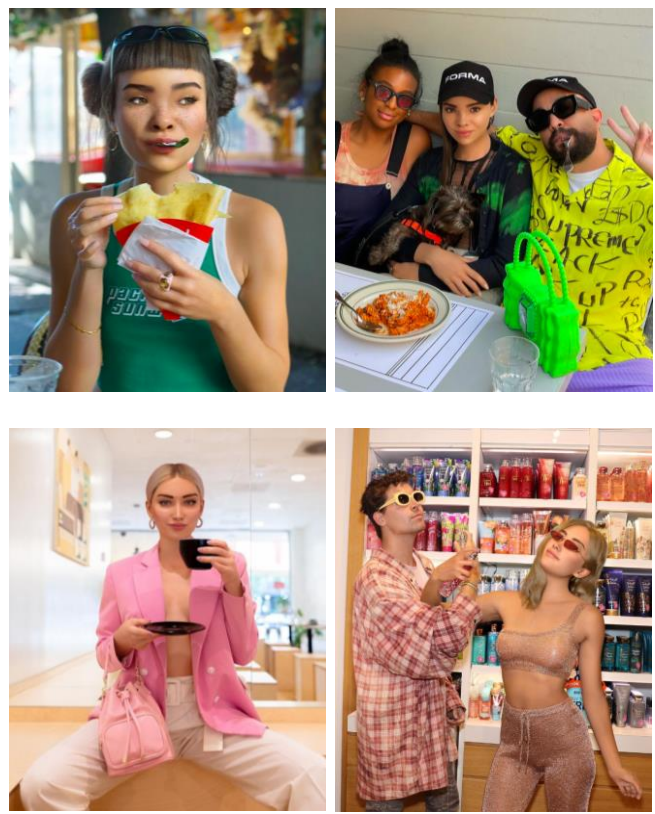

NOTE. — Lil Miquela (above) and Bermuda (below): alone vs. with companions.

This work makes four main contributions. First, we contribute to work on influencer marketing. While a growing body of research has examined influencer effectiveness (e.g., Leung et al. 2022c; Wies et al. 2022), less is known about virtual influencers. Further, while prior work has focused on influencer characteristics, language, or post disclosure (Hughes, Swaminathan, and Brooks 2019; Karagür et al. 2022), there has been less attention to visual content. Indeed, while visual content shapes online engagement (Li and Xie 2020; Overgoor et al. 2022; Hartmann et al. 2021), there has been relatively little work exploring what visual features matter. We fill this gap, examining a particular aspect of photos (i.e., whether there is someone else in them) and whether that increases posts' impact.

Second, we shed light on consumer interactions with non-human agents. From texting with chatbots to chatting with Alexa, non-human agents are playing an ever-greater role in marketplace interactions (Longoni, Bonezzi, and Morewedge 2019; Puntoni et al. 2021; Melzner, Bonezzi, and Meyvis 2022). But while avatars, chatbots, and other AI are clearly here to stay, relatively little is known about how such interaction partners shape consumer behavior. We demonstrate that appearing with others increases engagement with, and impact of, non-human agents because it makes them seem more like real people. Along the way we contribute to literature on anthropomorphism, showing that beyond characteristics of agents themselves (e.g., their name), other cues (e.g., being with others) shape inferences about how human they seem.

Third, we deepen understanding around consumer trust, and how to communicate it in online contexts. Trust is a fundamental part of social interactions, and shapes information seeking (Engeler and Barasz 2021), persuasion (Packard, Gershoff, and Wooten 2016), and engagement (Karagür et al. 2022). But what drives trust in the first place? We demonstrate that appearing with others increases trust in virtual influencers because it makes them seem more human.

Finally, from a substantive perspective, these findings have clear practical implications. Marketers and content creators want to boost engagement and drive choice. Our results demonstrate a simple, and easily implementable way to increase the impact of virtual influencer posts. Results of the field data, for example, suggests that pictures that include others receive 27% more engagement, or approximately 3,300 more likes or comments.

## NON-HUMAN AGENTS

Virtual agents have become integral part of marketplace interactions. Consumers buy from Alexa, chat with Siri, and get customer service from chatbots. Medical AI already diagnose heart disease and skin cancer (Hutson 2017; Haenssle et al. 2018) and conversational AI like ChatGPT promise to revolutionize everything from information search to content production.

Consistent with its increasing importance, consumer behavior scholars are beginning to study non-human agents and how they impact consumer behavior. Consumers are resistant to medical AI, for example, because they are concerned that it will be less able to account for their unique characteristics and circumstances (Longoni, Bonezzi, and Morewedge 2019). Algorithmic failures are generalized more than human failures because consumers assume that AI systems are more homogeneous (Longoni, Cian, and Kyung 2022). And when it comes to decisions that yield disparities (e.g., gender or race), however, people believe that algorithms are less biased than humans because they blindly applying rules and procedures irrespective of whom they are judging (Bonezzi and Ostinelli 2021).

But while a burgeoning stream of research has started to examine avatars, chatbots, and other AI, there has been less attention to a different type of non-human agents: virtual influencers.

# VIRTUAL INFLUENCERS

Other individuals are an important driver of marketing communications. Consumers know companies are trying to convince them to buy products and services, so they're less likely to trust traditional advertisements or other types of company produced content (Wies et al. 2022). Further, while word of mouth has always existed, the advent of online reviews, social media, and other digital content has made it easier to see what others think. Consequently, others' opinions have a huge impact on consumer choice (Moore and Lafreniere 2020; Liu, McFerran, and Haws 2020).

Building on this shift, marketers are increasingly relying on influencers to spread information and influence. Whether defined by what they do (e.g., fitness influencers), who they are (e.g., senior influencers), or where they go (e.g., travel influencers), people with large social media followings can impact trends and behaviors (Haenlein et al. 2020). Consistent with its importance, more than 75% of marketers now incentivize influencers to say positive things about products and services (Leung et al. 2022a).

That said, working with traditional influencers can be challenging. Unlike traditional paid or owned media, where brands completely control the communication (i.e., who says what, when), traditional influencers are harder to control. Brands can pick which influencer to work with, and suggest how they would like to see the brand portrayed, but they don't always have the final say about exactly what gets posted. Further, influencers have their own images to manage and may not perfectly execute the brand's vision. Finally, misbehavior or misalignment in their personal lives can spillover to damage the brand (Leung et al. 2022b). Tiger Woods' personal issues caused a 2% lost in market value for brands he represented (e.g., Nike and PepsiCo; Knittel and Stango 2014), for example, and while Ellen promoted Samsung phones via her infamous Oscar selfie, she later tweeted from her iPhone backstage.

Consequently, in response to desires for increased control, more and more companies are now shifting resources to virtual influencers (Miao et al. 2022; Schwarz 2022). Virtual influencers are avatars created by tech savvy companies using computer graphics. They are designed to look and behave like real influencers, having their own personalities, hobbies, and interests they post about (Leung et al. 2022b; see table 1 for examples). Given that creators have complete control over how virtual influencers look, and what is posted, virtual influencers provide a safer and easy way for brands to craft relevant messages (Open Influence 2022). Further, virtual influencers are digital beings, so their inherent connection with digital goods (e.g., NFTs or video game skins) makes them better spokespeople for metaverse-related things (e.g., Gen Z; Matthews 2022; McKinsey 2022). Approximately 75% of Gen Z follow at least one virtual influencer, and up to 40% have purchased a product virtual influencers promoted (Influencer Marketing Factory 2022).

**TABLE 1**: POPULAR VIRTUAL INFLUENCERS ON INSTAGRAM

| Virtual Influencer | Country | Description | Followers |
|---|---|---|---|
| Lil Miquela (@lilmiquela) | US | 19-year-old Brazilian American model and social activist. Testimonial for Prada, Samsung, and Dior. Singer with 440,000 monthly listeners on Spotify. | 3,041,852 |
| Lu Do Magalu (@magazineluiza) | Brazil | Shares reviews and software tips on various product releases (e.g., home appliance, clothing). | 6,011,784 |
| Imma (@imma.gram) | Japan | Tokyo-based testimonial for IKEA, Puma, and Porsche. Art museum and fashion passionate, with interest in environment, race, and gender issues. | 405,000 |
| Esther Olofsson (@esther.olofsson) | Sweden | Architecture, food, and styling addicted. Based in Rotterdam, writes stories and share pictures. | 43,200 |
| Bermuda (@bermudaisbae) | US | Musician and Trump supporter based in LA. Supports female entrepreneurs to pursue their business in the intersection of tech and beauty. | 266,100 |
| Blawko (@blawko22) | US | Self-proclaimed life lover. Based in LA, spends his free time venting on YouTube to his subscribers. | 139,000 |
| Shudu (@shudu.gram) | England | Digital fashion supermodel. Works with Cosmopolitan, Vogue, Balmain, and poses in premier and mystical shoot locations. | 236,000 |
| Ayayi (@ayayi.iiiii) | China | First Chinese meta-human, passionate of fashion and beauty. Partners Tiffany, MAC, and L'Oréal. | 12,000 |
| Rozy Oh (@rozy.gram) | South Korea | Loved by Gen Z, communicates daily lives through social media uploads and magazine photoshoots. | 145,000 |
| Kim Zulu (@kimzulu) | South Africa | First black female virtual influencer. Fashion and modelling passionate, brand ambassador for Puma. | 3,692 |

NOTE.— Info content is from virtual influencers' Instagram accounts, virtualhumans.org, and hypeauditor.com.

But while virtual influencers have a variety of benefits, their effectiveness depends on actually connecting with an audience. So, what might increase the impact of their posts?

## INFLUENCER EFFECTIVENESS

A growing literature has explored what shapes influencer effectiveness. Expertise, for example, boosts purchase (Jin and Phua 2014) and engagement (Hughes et al. 2019) because it signals confidence in what it is said. Disclosing that a post is paid increases word of mouth (Gerrath and Usrey 2021) and shares (Chen, Yan, and Smith 2022) because it makes influencers seem more authentic. Other work finds that factors like message positivity, influencer-brand fit, and follower count also shape engagement, choice, and persuasion (Karagür et al. 2022; Leung et al. 2022c; Wies et al. 2022).

But while some work has examined influencer effectiveness, less attention has been devoted to responses to virtual influencers. Further, it's not clear that the same factors apply. While expertise boosts the impact of regular influencers, for example, it is unclear what expertise means in the context of virtual influencers, as they can't experience the things they are endorsing (e.g., taste the food or feel the clothing). Similarly, while disclosing a post is paid may help regular influencers, given they aren't real, followers may already assume most of what virtual influencers are doing has some persuasive element.

## COMPANIONS AND TRUST

To begin to explore virtual influencers' impact, we focus on including others in images. Specifically, we suggest that showing the virtual influencer with others increases impact (i.e., engagement and persuasion) because it makes the influencer seem more human, and thus, more trustworthy.

Trust plays an important role in social interactions (Cohen and Isaac 2021). The more a source seems sincere, or motivated to provide accurate information, the more likely consumers are to follow their recommendations (Pornpitakpan 2004). Indeed, being seen as trustworthy makes consumers more willing to book an apartment (Ert and Fleischer 2019), buy wine (Packard et al. 2016), or invest money (Wilson and Eckely 2006).

Trust in virtual influencers, however, is relatively low (Open Influence 2020; Miao et al. 2022; Sands et al. 2022). Their account bios clearly state that they are fake, and they sometimes mention this in their posts, both of which should undermine trust and increase the salience of persuasion motives. Further, they can't have tried the product or service they are recommending (because they aren't real). Indeed, related work find that consumers trust avatars less than peers (De Visser et al. 2016).

That said, things that make avatars seem more human can increase trust (Tourè-Tillery and McGill 2015). Anthropomorphism is defined as seeming human in characteristics, intentions, emotions, beliefs, and mind (Aggarwal and McGill 2007), and having realistic traits can make non-human actors seem more human (Miao et al. 2022). Using a female voice, for example, makes virtual assistants seem more real (Wang et al. 2007), as does giving chatbots human names (e.g., "Jamie" rather than Spotifybot; Crolic et al. 2022).

We suggest that appearing with others should have similar effects. Being with others (e.g., friends or strangers) can influence consumption decisions (Belk 1975; Argo, Dahl and Machanda 2005; McFerran et al. 2010). But beyond how others' presence affects consumers *own* choices, we suggest it should impact how they perceive *others* (in this case, virtual influencers). Showing a virtual influencer hugging someone, hanging out with friends, or doing other things real humans do should make them seem more human. It should make them seem like they have friends, and are genuine social actors, which should make them seem more trustworthy. Consequently, appearing in a photo with others should increase trust because it makes virtual influencers seem more like real people.

This, in turn, should increase virtual influencers' impact. Virtual agents seeming human can have a host of positive consequences for choice, purchase, and persuasion (Holzwarth, Janiszewski, and Neumann 2006; Lin, Doong, and Eisingerich 2021; Tourè-Tillery and McGill 2015). Consequently, while consumers may naturally be wary of virtual influencers, factors that make them seem more human should increase trust, which should make consumers more willing to engage with, or be impacted by, whatever they post.

## THE CURRENT RESEARCH

A multimethod approach tests these possibilities. Study 1 provides a preliminary test in the field. We use text and image analysis to analyze almost 10,000 virtual influencer's Instagram posts, examining whether pictures that include a companion receive more engagement. We also include dozens of controls to account for potential sources of endogeneity. Ancillary analyses begin to test the hypothesized process, examining whether companion's effects disappear for human influencers (because the poster already seems real), when the post features a brand (because trust is in greater doubt to begin with), or when the campaign is encouraging purchase (because trust is in greater doubt to begin with).

To directly test companion's causal impact, and further explore the hypothesized underlying process, the next two studies use experiments. Study 2 manipulates companion presence and examines whether it increases engagement. Further, it tests whether companion presence boosts engagement because it makes the virtual influencer seem more real, which makes them seem more trustworthy. Study 3 further tests the hypothesized process through moderation, examining whether the effects extend to likelihood of choice. If trustworthiness is driving the effects of companion presence, as we suggest, then its effects should be mitigated when influencer realness is not in doubt (i.e., when the poster is human).

## STUDY 1: ALMOST 10,000 VIRTUAL INFLUENCER POSTS

To provide a preliminary test of the link between companions and engagement, study 1 looks to the field. We use automated image analysis to analyze almost 10,000 virtual influencers' Instagram posts, testing whether, controlling for a range of other factors, posts that include a companion receive more engagement.

Method

A leading influencer marketing agency provided a sample of 9,766 Instagram posts from 28 virtual influencers. The sample includes influencers from various countries (e.g., US, Japan, and UK) that shared at least one sponsored post between October 14, 2014, and August 9, 2022. The posts cover products and services and include a range of industries (e.g., beauty, travel, and gaming). Most posts were in English (94%), and Upwork professionals translated the rest into English to enable the use of established linguistic dictionaries.

We used image mining to detect the presence of a companion in any of the 19,158 pictures included in the post (posts can have more than one picture). Following prior work (Li and Xie 2020; Villarroel Ordenes and Silipo 2021), we used the Google Cloud Vision API's "face detection" service to detect the number of faces within a picture. A random sample of 100 pictures found 100% accuracy in detecting virtual influencers' faces. Any post where at least one picture featured more than one face was flagged as potentially involving a companion. Manual checks ensured that each photo actually included the influencer (rather than just multiple other people or a meme). The data set also includes 1,047 videos for which companion presence was manually annotated by a hypothesis blind research assistant.

Following prior research (e.g., Herhausen et al. 2019), engagement was measured as the number of likes and comments posts received. On average, posts received 18,292 likes (SD = 46,659, ranging from 15 to 2,195,403) and 305 comments (SD = 1,221, ranging from 0 to 58,478). Descriptive statistics and correlations are in web appendix (table A1).

Finally, we examined the relationship between companion and engagement. Given the dependent variable is count and overdispersed ($p < .001$, likelihood ratio test), a negative binomial regression was used (results are also robust to OLS regression with log transformed DV, see robustness checks below). Given the different variables do not share similar scales, all continuous variables were standardized (z-scored). Unstandardized results are the same in sign and significance. To provide directly interpretable measure of the variables' effect size, we express coefficients as incident rate ratios (IRRs), so greater than 1 means positive impact and below 1 means negative impact.

Results

Consistent with our theorizing, when virtual influencers pictures included a companion, their posts received greater engagement ($IRR = 1.699$; $SE = .068$, $t = 13.24$; $p < .001$; table 2, model 1).

*Control variables.* While this initial result is intriguing, one could wonder whether it is driven by other factors. Consequently, we control for a range of alternative explanations, including aspects of the influencer, image, text, and time effects.

*Aspects of the Influencer.* Rather than the presence of a companion, one could wonder whether the results are driven by something about the virtual influencer posting the content. If an influencer has more followers, for example, more people may see their posts, which should lead to more likes and comments (Wies et al. 2022). Similarly, some influencers have verified accounts, which may lead to higher engagement (Valsesia, Proserpio, and Nunes 2020). Consequently, we control for both aspects (replacing both time-invariant factors with influencer fixed effects finds similar results). Finally, while it is not exactly about the influencer per se, when influencers post more frequently, followers may infer that the information is fresh and up to-date, which may increase engagement (Leung et al. 2022c). Consequently, we control for the total number of posts shared.

*Aspects of the Image.* Beyond the influencer posting, we also control for aspects of the image itself. First, images and videos might encourage different levels of engagement (Tellis et al. 2019), so we control for post type (image or video). Second, color dominance and saturation enhance viewers' attention and can increase engagement (Li and Xie 2020), so we measured these features and included them as controls.[2] Third, visual complexity, like object count (Pieters, Wedel, and Batra 2010) can make images harder to understand, which can affect engagement (Overgoor et al. 2021), so, we used the Google's API to detect the distinct objects within an image (e.g., smartphone or lamp) and include object count as a control. Fourth, we also used the API to control for the emotional state of faces that appeared (i.e., anger, joy, and surprise). The service assigns to each face a score for each emotion on a 5-point scale, posts can feature multiple images, and each image contains multiple faces, so we averaged emotion scores across them.

---

[2] Following prior work (Villarroel Ordenes et al. 2019), we extracted the first image that appeared in each video to compute color dominance and saturation (averaging scores across images of the same post) using the Python's Image module from PIL.

*Aspects of the Text.* Beyond who posts, or the post's image, the accompanying text may also impact engagement. Consequently, we control for aspects of the text. First, one could wonder whether some topics or theme get more engagement, and this, rather than the presence of a companion, is driving the effect. To address this, we control for the topics discussed in each post. We used Empath (Fast, Chen, and Bernstein 2016), a tool that analyzes text across 194 built-in, pre-validated categories (e.g., traveling, technology, and cooking),[3] and to reduce dimensionality, we used factor analysis with varimax rotation. This identified 70 overarching topics.[4] Second, virtual influencers often ask questions to prompt dialogues with followers, which can increase engagement (Villarroel Ordenes et al. 2019), so we control for the number of questions within the message. Third, given that hashtags can boost the number of people who see a post, and thus engagement, we control for the number of hashtags. Fourth, longer posts may convey more information, which could impact engagement, so we control for the length of the post (in words). Fifth, the number of emojis may impact engagement (Luangrath, Xu, and Wang 2022), so we control for it. Sixth, some posts mention brands or the accounts of people that appear in the photos, which could boost visibility and thus engagement, so we control for the number of mentions. Seventh, arousal and valence can affect engagement (Berger and Milkman 2012), so we control for both aspects using the Mohammad's (2018) VAD (Valence, Arousal, Dominance) lexicon. Eighth, easy-to-read posts might be easier to process, which may increase engagement (Pancer et al. 2019), so we control for text complexity using Flesch–Kincaid (Flesch 1948). Ninth, a more concrete language can increase engagement because it signals direct experience (Packard and Berger 2021), so we control for linguistic concreteness using Paetzold and Specia's (2016) ratings. Tenth, familiar words are easier to be processed and thus can boost engagement more (Pancer et al. 2019), so we control for it using Paetzold and Specia's (2016) familiarity scores. Eleventh, greater extremity can increase engagement because it makes posts seem more helpful (Rocklage and Fazio 2020), so we control for extremity using the Rocklage, Rucker, and Nordgren's (2018) Evaluative Lexicon 2.0. Twelfth, social words (i.e., "friends" or "together" Lasaleta, Sedikides, and Vohs 2014) can boost some forms of engagement because it makes the speaker seem friendlier (Park et al. 2021), so we control for them using the Linguistic Inquiry and Word Count's (Pennebaker et al. 2015) "social" measure. Thirteenth, some posts are sponsored, which could help or hurt engagement (Leung et al. 2022c), so we control for this factor as well. Finally, posts also sometimes offer incentives (e.g., free gifts or discount codes) for followers who leave a comment or tag other people (Hughes et al. 2019). Such incentives might increase engagement, so we include a dummy variable to account for whether a post contains such a sales promotion, according to Jalali and Papatla's (2019) list of promotional words (i.e., chance, commercial, free, gift, giveaway, promo, win, and sale).

---

[3] We did not use Latent Dirichlet Allocation (LDA) because it provides weak coherence and efficacy on short messy text like social media posts (Mehrotra et al. 2013).
[4] We used varimax rotation to minimize the number of variables that have high loadings on each factor.

*Time Effects.* Given certain times of day, weekdays rather than weekends, or time of year may get more engagement, we control for when content was posted using fixed effects for time of day, weekdays, month, and year.

*Results Including Controls.* Even after accounting for all these controls, however, posts that included a companion still received more engagement (*IRR* = 1.273; *SE* = .049; *t* = 6.31; *p* < .001; table 2, model 2). Such posts are associated with a 27.3% increase in engagement and receive 3,306 more likes or comments, on average.[5]

---

[5] The direction and size of the effect is similar if the companion is a real person (*IRR* = 1.277) or virtual one (*IRR* = 1.164).

**TABLE 2:** COMPANION AND ENGAGEMENT

|  | (1) Base Model | (2) With Controls |
|---|---|---|
| **Companion Presence** | **1.699*** (.068)** | **1.273*** (.049)** |
|  |  |  |
| Controls |  |  |
| *Influencer* |  |  |
| # of Posts |  | 1.517*** (.050) |
| Follower Count |  | .924** (.027) |
| if Verified |  | 2.120*** (.099) |
| *Image* |  |  |
| Image (vs. Video) |  | 1.057 (.045) |
| Color Dominance |  | .881*** (.009) |
| Color Saturation |  | .944*** (.013) |
| Visual Complexity |  | .925*** (.013) |
| Anger |  | 1.012 (.073) |
| Joy |  | 1.116*** (.021) |
| Sorrow |  | 1.180** (.065) |
| Surprise |  | 1.021 (.048) |
| *Text* |  |  |
| Topic discussed |  | Included |
| # of Questions |  | 1.081*** (.017) |
| # of Hashtags |  | 1.104*** (.022) |
| Wordcount |  | .843*** (.039) |
| # of Emojis |  | .988 (.015) |
| # of Mentions |  | .973* (.013) |
| Arousal |  | 1.001 (.016) |
| Valence |  | .926*** (.015) |
| Complexity |  | .873** (.037) |
| Concreteness |  | 1.205*** (.021) |
| Familiarity |  | 1.061*** (.016) |
| Extremity |  | 1.022 (.015) |
| Social Words |  | 1.485*** (.029) |
| Sponsored Post |  | .972 (.053) |
| Promotional Post |  | .749*** (.029) |
| *Time Fixed Effects* |  | Included |
| Log likelihood | –103,098 | –100,930 |
| N | 9,766 | 9,766 |

\* *p* < .05, \*\* *p* <. 01, \*\*\* *p* < .001. Estimates are IRRs. Standard errors are in parentheses.

Robustness

We also ran a number of additional robustness tests. First, one might argue that inclusion of a companion in an image might not be random. There might be observable or unobservable factors that drive propensity to have companions in posts. To attempt to address such endogeneity concern, we use propensity score matching (Rosenbaum and Rubin 1983) to "adjust" for differences in the treatment and control group which may bias inferences about the treatment (see the web appendix for details). In our case, the propensity score is the predicted probability that a post receives the treatment (i.e., includes a companion in the image) conditional on the value of covariates. Following Li and Xie (2020), we also used the percentage of posts that contain companion in the influencer account history as an instrument. We adopted a 1:1 nearest-neighbor matching algorithm without replacement and a caliper of .01 to match a post including companion with posts without, with the closest propensity score (see table A2 for statistics of variables before and after matching). Even accounting for this endogeneity correction, however, results remain the same. Posts that include others in the image receive more engagement ($IRR = 1.130$; $SE = .050$; $t = 2.74$; $p = .006$; table A3, column 1).

Second, one could wonder whether the results are somehow driven by the modeling approach used. In particular, one could argue the ranges of the data and extreme values (i.e., engagement ranges from 2 to 2,252,664) might make the use of count distributions less appropriate. To address this, we re-run the analysis using an ordinary least squares regression with a log-transformed dependent variable. Results remain the same ($b = .117$; $SE = .042$; $t = 2.76$; $p = .006$; table A3, column 2).

Third, while the combination of likes and comments were used to measure engagement, results remain similar when each is examined separately (likes: $IRR = 1.277$; $SE = .049$; $t = 6.32$; $p < .001$; comments: $IRR = 1.058$; $SE = .033$; $t = 1.81$; $p = .071$; table A3, columns 3 and 4, respectively).

Fourth, one could argue that the number of likes and comments virtual influencers receive on their last post might influence the visibility of the next post. So, we controlled for potential carryover effects by including the lagged dependent variable in the predictor set. Results remain the same ($IRR = 1.143$; $SE = .041$; $t = 3.69$; $p < .001$).

Fifth, one could wonder whether rather than being about the presence of a companion, maybe the effects are driven by no individuals in an image hurting engagement. But this does not seem to be the case. Posts which include a companion receive more engagement than both posts containing no one ($IRR = 1.291$; $SE = .093$; $t = 7.91$; $p < .001$) or just one individual (i.e., just the influencer themselves, $IRR = 1.262$; $SE = .049$; $t = 6.02$; $p < .001$).

Sixth, perhaps posts with companions are more likely to depict active product consumption (e.g., using a lipstick or eating at a restaurant), and this drove the effect. To test this possibility, a research assistant annotated a random sample of 500 posts in terms of product use (dummy-coded). Companion presence and product use were not related ($r = .001$), however, casting the doubt on the possibility that product use is driving things.

Finally, one might wonder if, rather than companion presence, the effects are driven by the specific objects virtual influencers display in pictures. Things like moisturizers, hamburgers, or beaches, for example, could attract more engagement than clippers, cabbages, or bridges. Consequently, we used Google's API to extract all objects depicted in pictures and include them as controls in the model. Even after accounting for 2,005 object controls, however, the effect of companion persisted ($IRR = 1.232$; $SE = .048$; $t = 5.31$; $p < .001$).

Discussion

Study 1 provides initial support for our theorizing in the field. Text and image analysis of almost 10,000 virtual influencers' posts demonstrates that posts featuring a companion receive more than 27% more engagement. Including various controls casts doubt on alternative explanations and speaks to the generalizability of the effects.

**EXPLORING THE UNDERLYING PROCESS**

While the main point of study 1 was to test the relationship between companion presence and engagement in the field, we also use this data to begin to test the hypothesized process behind the effects (studies 2 and 3 do this more directly). Specifically, to test whether the effects are driven by the humanizing impact of others' presence, and the effect it has on perceived trust, we examine the moderating role of (1) influencer type (i.e., virtual vs. human), (2) brand presence, and (3) post intent.

Moderating Role of Influencer Type

If the presence of a companion increases engagement by making consumers feel that the virtual influencer is real and thus more trustworthy, as we suggest, then its effect should be mitigated in situations where the source's realness is not in doubt. To test this possibility, we examined the moderating role of influencer type by comparing virtual and human influencers.

We worked with the agency to acquire a sample of 22,411 Instagram posts from 1,429 human influencers. Posts include 38,607 images and cover similar industries (see table A4 for descriptive statistics and correlations). Given there might be systematic differences between virtual and real influencers (e.g., post frequency or tendency to share videos), propensity score matching was used to identify real influencers who were as similar as possible to virtual influencers in terms of traits and posting behavior (i.e., text, image, and posting time; see web appendix for details). In the matching equation predicting whether the influencer was virtual or not, we used all the covariates of our original engagement model with controls. The resulting matched sample included 19,532 posts, half from virtual influencers and half from human influencers (see table A5 for statistics of variables before and after matching).

As predicted, results revealed a significant companion × poster type (i.e., virtual vs. human) interaction ($IRR = 1.110$; $SE = .048$; $t = 2.20$; $p = .027$; table A6). Consistent with our suggestion about the humanizing impact of including others in a post, while companion boosted engagement with virtual influencers ($IRR = 1.144$; $SE = .045$; $t = 3.43$; $p = .001$), its effect was mitigated for human influencers ($IRR = 1.026$; $SE = .019$; $t = 1.38$; $p = .169$). Study 3 further tests this moderation using an experiment.

This result also casts doubt on various alternative explanation. Rather than being driven by anthropomorphism and trust, as we suggest, one could wonder whether the effects are driven by something more general, such as liking, popularity, or social proof. Maybe the presence of someone else in the photo makes it seem like the poster is more likable, for example, or more popular, and these aspects are driving the results. But while such explanations are plausible, the moderation by poster type casts doubt on these possibilities. Rather being just generally beneficial, consistent with our theorizing about the role of anthropomorphism and trust, results suggest that there is something specific about virtual influencers (compared to human influencers), that makes others' presence particularly beneficial for them. Study 3 further tests these alternatives.

Moderating Role of Brand Presence

If companion presence increases engagement because it makes consumers believe the virtual influencer is more trustworthy, as we suggest, then the effect should be stronger in situations where trust is more in question. To test this possibility, we explored the moderating role of brand presence.

Research on persuasion knowledge notes that it is activated by the presence of relevant cues (Karagür et al. 2022). In this case, some posts are organic (e.g., the influencer talks about the weather) while others are more clearly focused on brands. Posts that mention brands should be more likely to make observers wonder whether the poster was paid to post, and thus more skeptical of their trustworthiness. To measure brand presence, we used a list of the top 1,000 brands (https://www.comparably.com/brands/top-1000-brands) to identify posts which mention a brand in the text (i.e., tag (@), hashtag (#) or simple mention), and Google Cloud Vision API's "logo detection" function to detect the presence of a brand logo in the image. These were combined to indicate brand presence in the post (dummy coded: 1 if present, 0 otherwise).

A negative main effect of brand presence on engagement ($IRR = .676$; $SE = .021$; $t = -12.75$; $p < .001$) was qualified by the predicted companion × brand presence interaction ($IRR = 1.368$; $SE = .102$; $t = 4.19$; $p < .001$). Specifically, consistent with the notion that the effects are driven by trustworthiness, the presence of a companion was more beneficial when influencer trustworthiness was in greater doubt (i.e., when the post mentioned a brand).

Moderating Role of Post Intent

We also explored the moderating role of post intent. While some posts try to boost awareness (e.g., encourage followers to learn more about a product or service), or raise interest, others are designed to persuade (i.e., to encourage purchase). Given their focus, these types of posts should be particularly likely to raise questions about poster trustworthiness (i.e., whether or not they were paid to post; Hughes et al. 2019). Consequently, if our theorizing is correct, companion presence should have larger an effect when posts are encouraging purchase.

To test this possibility, we created a dictionary of 53 words indicating a purchase-oriented intent of the post (e.g., "sale," "gift," or "buy"; see web appendix for approach used and full list of words). Then, we took all posts featuring a brand, and measured whether the language was aimed at encouraging purchase (i.e., if at least one of the purchase-oriented words appeared in the post).

Results indicate that a negative main effect of purchase intent ($IRR = .820$; $SE = .017$; $t = –9.48$; $p < .001$) was qualified by the predicted companion × purchase interaction ($IRR = 1.085$; $SE = .042$; $t = 2.08$; $p = .037$). Consistent with our theorizing about the role of trust, companion presence had a larger effect when influencer trustworthiness was more in doubt (i.e., when the post was focused more on purchase).

Taken together, these three moderation tests provide further evidence for the hypothesized underlying process. Studies 2 and 3 use experiments to further test the process.

## STUDY 2: MANIPULATING COMPANION PRESENCE

Study 2 has two main goals. First, while the results of study 1 are consistent with the notion that companion presence increases engagement with virtual influencers' posts, one could still wonder whether the relationship is truly causal. Controlling for a variety of factors cast doubt on alternative explanations, but an even stronger test would be to manipulate companion presence and examine whether it increases engagement. Study 2 does this.

Second, building on the study 1 moderation analyses, study 2 further tests the hypothesized mechanism. We have suggested that companion presence increases engagement because it makes consumers feel like the virtual influencer is real. This, in turn, increases perceived trust, which increases engagement. Study 2 tests this sequential process.

Method

Participants ($N = 140$; Prolific) were randomly assigned to condition in a between-subjects design. See web appendix for materials, exclusions, demographics, and power analyses for all experiments. Everyone was shown an Instagram post endorsing a bar. They were told it was posted by a virtual influencer named Isla Robin and that "virtual influencers are computed-generated influencers that look human but are not. Further, they post on social media as real influencers do."

The only difference between condition was companion presence. In the control condition, only Isla Robin appeared in the image, but in the companion condition, another person appeared with her.

Next, we collected process measures. Participants rated their perceptions of the virtual influencer's anthropomorphism using a 9-item scale adapted from Crolic et al. (2022, e.g., "to what extent the virtual influencer: came alive (like a person) in your mind, has some humanlike qualities, or seems like a person," 1 = not at all, 7 = to a great extent; α = .96). We also measured how trustworthy the influencer seemed using a 2-item scale adapted from Lin et al. (2021, "trustworthy" and "honest", $r = .79$)

Then, we measured the dependent variable. Participants were asked how likely they would be to engage with the post (i.e., like it or comment on it; Valsesia et al. 2020, 1 = not at all likely, 7 = very likely).

Finally, participants completed two attention checks and demographics.

Results

*Engagement.* As predicted, and consistent with study 1, including someone else in the image increased engagement ($M_{companion}$ = 2.33 vs $M_{control}$ = 1.62, $F(1, 138)$ = 6.37, $p$ = .013, $\eta^2$ = .044).

*Anthropomorphism.* Further, consistent with our theorizing, including someone else in the image also made participants think that the virtual influencer was more real ($M_{companion}$ = 3.70, vs $M_{control}$ = 3.15, $F(1, 138)$ = 4.62, $p$ = .033, $\eta^2$ = .032).

*Trust.* In addition, including someone else in the image also made the virtual influencer seem more trustworthy ($M_{companion}$ = 3.31 vs $M_{control}$ = 2.71, $F(1, 138)$ = 7.97, $p$ = .005, $\eta^2$ = .055).

*Mediation.* Finally, serial mediation (PROCESS model 6; Hayes 2018) supported the hypothesized underlying process ($b$ = 0.13, 95% CI = 0.01, 0.28). Including someone else in the image made consumers think that the virtual influencer was more real ($b$ = 0.55, $SE$ = 0.26, $t$ = 2.15, $p$ = .033), which made the influencer seem more trustworthy ($b$ = 0.40, $SE$ = 0.06, $t$ = 6.56, $p$ < .001), which increased engagement ($b$ = 0.57, $SE$ = 0.11, $t$ = 5.13, $p$ < .001). Including these mediators led the direct effect to be reduced to non-significance ($b$ = 0.27, 95% CI = –0.22, 0.76), indicating full mediation.

Discussion

Study 2 provides direct causal evidence that companion presence increases engagement and underscores the hypothesized process behind the effect. First, consistent with study 1, companion presence increased engagement. Including another person in an image with a virtual influencer increased observers' willingness to engage with the post.

Second, consistent with our theorizing, serial mediation demonstrates that these effects were driven by anthropomorphism and perceived trust. Including a companion in an image made virtual influencers seem like they were more real, which made them seem more trustworthy, which boosted engagement.

## STUDY 3: PROCESS BY MODERATION

Study 3 has three main goals. First, it further tests the hypothesized process through both mediation and moderation. If companion presence increases engagement by making it seem like virtual influencers are more real and thus more trustworthy, as we suggest, then the effect should be mitigated when the realness of the source is not in question. To test this possibility, in addition to manipulating companion presence, we manipulate influencer type (i.e., virtual vs. human, similar to the ancillary analysis in study 1). Said another, while study 2 measures anthropomorphism perceptions, study 3 manipulates them. If our theorizing is correct, the effect of companion presence on engagement should be mitigated when the poster is clearly human (as shown in study 1).

Second, we examine an additional dependent variable. While engagement is correlated with sales (Goh, Heng, and Lin 2013; Rishika et al. 2013; Kumar et al. 2016), one could wonder whether the observed effects extend to likelihood of choosing the endorsed product. We test this possibility.

Third, we test alternative explanations (see study discussion for more detail).

Method

Participants ($N = 280$) were randomly assigned to condition in a 2 (companion: yes vs. control) × 2 (influencer type: virtual vs. human) between-subjects design.

In addition to manipulating companion presence (using the design from study 2), we also manipulated influencer type. In the virtual [human] condition, participants were told that the poster was "a virtual [human] influencer on Instagram." Participants in the virtual influencer condition also read the definition of a virtual influencer from study 2.

In addition to measuring engagement and trustworthiness using the measures from study 2, we also examined choice likelihood. Participants were asked how likely they would be to choose the endorsed bar in the future (1 = not at all; 7 = very).

Finally, participants completed ancillary measures to test alternative explanations (i.e., popularity and likeability; see more detail below), two attention checks, and demographics.

Results

*Engagement.* A main effect of influencer type ($F(1, 278) = 4.20$, $p = .041$, $\eta^2 = .015$) was qualified by the predicted companion × influencer type interaction ($F(3, 276) = 3.70$, $p = .031$, $\eta^2_p = .017$). Consistent with the prior studies, when the influencer was virtual, companion presence increased engagement ($M_{companion} = 2.36$ vs $M_{control} = 1.62$, $F(1, 131) = 8.19$, $p = .005$, $\eta^2 = .059$). When the influencer was human, however, as expected and consistent with Study 1, this effect was mitigated ($M_{companion} = 2.34$ vs $M_{control} = 2.46$, $F(1, 145) = 0.17$, $p = .679$, $\eta^2 = .001$).

*Choice.* Effects were similar for likelihood of choice. A main effect of influencer type ($F(1, 278) = 4.95$, $p = .027$, $\eta^2 = .017$) was qualified by the predicted companion × influencer type interaction ($F(3, 276) = 3.56$, $p = .043$, $\eta^2_p = .015$; figure 2). Consistent with our theorizing, companion presence made people more likely to choose the endorsed bar when the influencer was virtual ($M_{companion} = 3.74$ vs $M_{control} = 3.12$, $F(1, 131) = 6.17$, $p = .014$, $\eta^2 = .045$), but this effect was mitigated when the influencer was human ($M_{companion} = 3.79$ vs $M_{control} = 3.91$, $F(1, 145) = 0.22$, $p = .640$, $\eta^2 = .001$).

**FIGURE 2:** MODERATION BY INFLUENCER TYPE (LIKELIHOOD OF CHOICE)

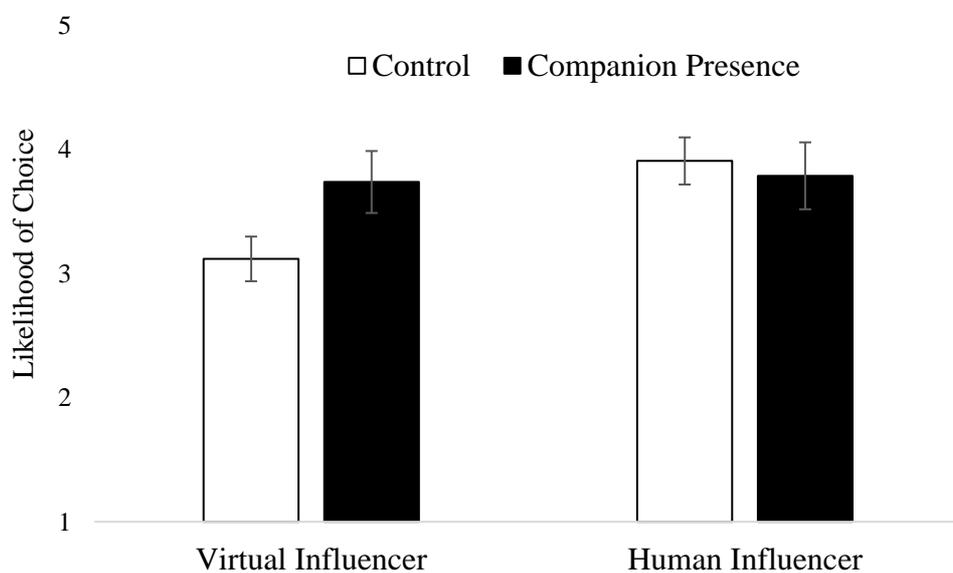

*Trust.* Main effects of both influencer type ($F(1, 275) = 28.48$, $p < .001$, $\eta^2 = .093$) and companion ($F(1, 275) = 6.03$, $p = .015$, $\eta^2 = .021$) were qualified by the predicted companion × influencer type interaction ($F(3, 276) = 14.67$, $p = .005$, $\eta^2_p = .028$; figure 3). Consistent with our theorizing, companion presence increased trust in virtual influencers ($M_{companion} = 3.58$ vs $M_{control} = 2.78$, $F(1, 131) = 12.74$, $p = .001$, $\eta^2 = .089$), but this effect was mitigated when the influencer was human ($M_{companion} = 3.97$ vs $M_{control} = 3.99$, $F(1, 145) = 0.01$, $p = .920$, $\eta^2 = .001$).

Looked at another way, while control participants trusted the influencer less when they were virtual ($M_{virtual}$ = 2.78 vs $M_{human}$ = 3.99, $F(1, 135) = 43.30$, $p < .001$, $\eta^2 = .243$), including a companion in the image increased trust almost to the level of human influencers ($M_{virtual}$ = 3.58 vs $M_{human}$ = 3.97, $F(1, 141) = 3.00$, $p = .086$, $\eta^2 = .021$).

**FIGURE 3:** MODERATION BY INFLUENCER TYPE (TRUST)

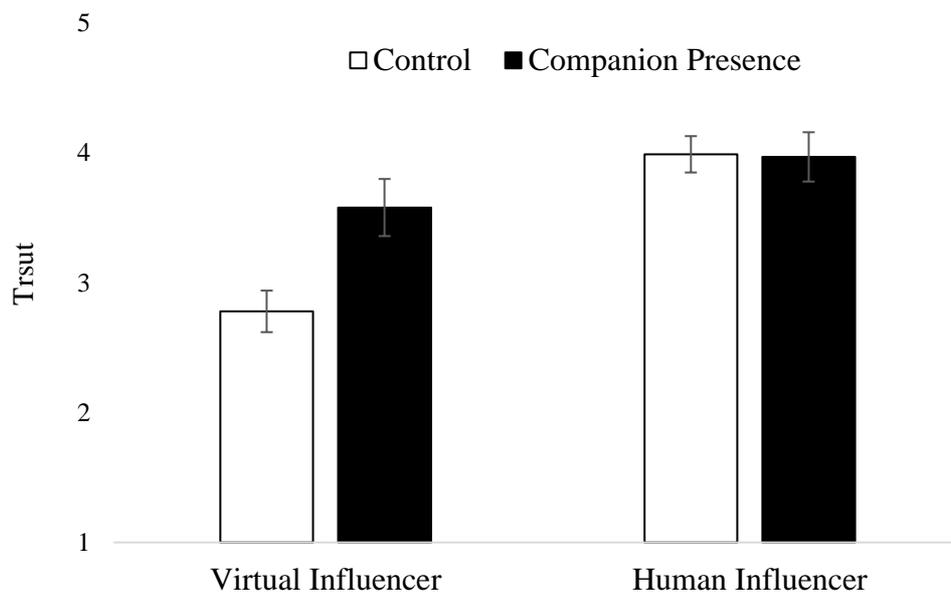

*Moderated Mediation.* Finally, moderated mediation (PROCESS model 8; Hayes 2018), incorporating influencer type as a moderator of companion's effect on trust, found significant moderated mediation on both engagement ($b = 0.59$, 95% CI = 0.16, 1.06) and likelihood of choice ($b = 0.52$, 95% CI = 0.16, 0.94).

As expected, in the virtual influencer condition, the effect of companion presence was driven by trust (engagement: $b = 0.58$, 95% CI = 0.23, 0.96; likelihood of choice: $b = 0.51$, 95% CI = 0.22, 0.86). Companion presence made the virtual influencer seem more trustworthy ($b = 0.80$, $SE = 0.21$, $t = 3.78$, $p < .001$), which increased both engagement ($b = 0.72$, $SE = 0.06$, $t = 11.34$, $p < .001$) and likelihood of choice ($b = 0.64$, $SE = 0.06$, $t = 10.38$,

$p < .001$). When the poster was human, however, trust was already high and thus companion presence no longer impacted trust ($b = –0.02$, $SE = 0.20$, $t = –0.09$, $p = .924$), and the mediation was no longer significant (engagement: $b = –0.01$, 95% CI = –0.28, 0.25; likelihood of choice: $b = –0.01$, 95% CI = –0.25, 0.22). Including the mediator in the model reduces the direct effect to a non-significance (engagement: $b = 0.01$, 95% CI = –0.31, 0.34; likelihood of choice: $b = –0.01$, 95% CI = –0.32, 0.31), indicating full mediation.

Discussion

Study 3 underscores the hypothesized underlying role of trust through both mediation and moderation. First, consistent with the prior studies, including someone else in an image increased engagement with virtual influencer's posts because it made consumers think the virtual influencer was more trustworthy. Second, consistent with the notion that these effects are driven by trust, however, when the influencer was seen as human, and thus more trustworthy to begin with, the effect disappeared.

Third, results demonstrate that these effects extend to likelihood of choosing the endorsed product. Consumers were more willing to choose the bar the virtual influencer endorsed when they appeared with another person in the photo.

*Alternative explanations*. Ancillary analyses also cast doubt on a number of alternative explanations. Given that companion presence was only expected to matter when the poster was a virtual influencer, we focus analyses there.

First, as noted in study 1, maybe others' presence makes the influencer seem more popular and that drove the effects. To test this possibility, we asked participants how popular they thought the influencer is (1 = not at all; 7 = very much). Popularity did not vary by condition ($F(1, 131) = 0.94$, $p = .334$, $\eta^2 = .007$), however, casting doubt on this alternative.

Second, as noted in study 1, maybe companions made the influencer seem more likeable, and that drove the effect. To test this possibility, we adapted Holzwarth et al.' (2006) two-item measure of likeability (unlikeable – likeable; unfriendly – friendly; $r = .77$). Although likeability was significantly higher in the companion condition ($M_{companion} = 3.85$ vs $M_{control} = 3.39$, $F(1, 131) = 4.37$, $p = .038$, $\eta^2 = .032$), it did not mediate the effects on either engagement ($b = 0.15$, 95% CI = –0.01, 0.37) or choice ($b = 0.10$, 95% CI = –0.01, 0.30).

Finally, as noted in study 1, one could wonder whether the results are somehow driven by social proof. Having more people in the photo at the bar, for example, could suggest more people like it, and thus it is better or more worth going to. Note, however, that this alternative has trouble explaining the companion's interaction with influencer type. While more people in a photo might be more beneficial, the number of people in the photo in the companion presence condition was the same in the human and virtual influencer conditions. Consequently, the fact that others' presence only boosted engagement and choice likelihood in the virtual influencer condition suggests it is something specific to that condition that is driving the effect.

Taken together, these analyses cast doubt on the possibility that popularity, likeability, or social proof, are driving the effects.

## GENERAL DISCUSSION

Influencers have become a powerful marketing tool. Consumers buy clothes influencers wear and visit restaurants influencers suggest. But while virtual influencers have become an emerging way companies can diffuse information and influence, less is known about what makes some virtual influencer posts more impactful than others.

This work explores a simple and common cue, the photos virtual influencers share. Specifically, a multimethod investigation, combining field data with controlled experiments, demonstrates the benefit of including others in photos.

First, automated image and text analysis of almost 10,000 Instagram posts demonstrates that posts which show virtual influencers with others receive more engagement (study 1). Follow up experiments manipulating others' presence provide direct causal evidence of these effects (studies 2 and 3). Including others in photos makes consumers more willing to like and/or comment on the post (studies 2 and 3) and choose the offering discussed (study 3).

Second, the results highlight the underlying role of trust and anthropomorphism in driving these effects. The presence of others increases engagement because it makes virtual influencers seem more like real people, which makes them seem more trustworthy (study 2). Further, consistent with the notion that the effects are driven by trust, they are stronger when trust is more in question (i.e., when the influencer is talking about a product or using language more related to buying a product, study 1 ancillary analyses) and mitigated when the source's realness is in less doubt (i.e., when the influencer is human, study 3 and study 1 ancillary analysis).

Third, finding consistent results across tens of thousands of real influencer posts and controlled experiments speaks to the generalizability of the effect. Similarly, demonstrating the underlying process through both mediation and moderation underscores its important role. Finally, including around 100 controls in the field (e.g., aspects of the influencer, text, visuals, and other features) and testing popularity, social proof, and likeability in the experimental work, help casts doubt on alternatives and underscores the effects' robustness.

Contributions and Implications

This research makes several contributions. First, it deepens understanding around influencers and influence. While a burgeoning stream of research has begun to examine the effectiveness of human influencer (Leung et al. 2022c; Wies et al. 2022), less is known about virtual influencers. Given more and more companies are relying on virtual influencers to achieve marketing goals, this area deserves more attention.

Second, we demonstrate the power of visual content. Despite the rapid proliferation of images and video in social media, few studies have focused on it (e.g., Liu, Dzyabura, and Mizik 2020; Hartmann et al. 2021). Indeed, while past work on influencers has examined things like language, disclosure, and poster characteristics (e.g., Hughes et al. 2019; Valsesia et al. 2020; Karagür et al. 2022), there has been less attention to characteristics of the pictures or images influencers post (Abell and Biswas 2022). We demonstrate that the presence of a companion in photos can have an important effect. Further, given that visual features play a key role in social media's impact (Li and Xie 2020), hopefully future work will examine them in greater detail.

Third, more broadly, this research contributes to literature about non-human agents. A growing body of work has begun to examine how avatars, chatbots, and others AI shape consumer behavior (Longoni et al. 2019; Puntoni et al. 2021; Caldario, Longoni, and Morewedge 2021; Longoni and Cian 2022; Bonezzi, Ostinelli, and Melzner 2022). Further, whether through voice (Wang et al. 2007), or name (Crolic et al. 2022), seeming like real people can significantly improve consumer attitudes and intentions. We add to these studies by demonstrating that, beyond characteristics of agent themselves, other cues (i.e., what they post) shapes inferences about how human they seem. Given that companies are heavily investing in non-human agents to engage and serve their customers better, this area is ripe for further attention.

Fourth, we advance knowledge on what drives consumer trust in online communication. While trust shapes almost every consumer choice (e.g., Packard et al. 2016), it has been relatively understudied in online contexts (Bleier and Eisenbeiss 2015; Bleier, Harmeling, and Palmatier 2019; Grewal and Stephen 2019; Karagür et al. 2022). Our research provides insights into the role of companion's presence, showing how it influences how real, and thus trustworthy, non-human agents seem.

Finally, our findings have important implications for practice. Many notable brands (e.g., Samsung, Prada, and Alibaba) have turned to virtual influencers (Bringè 2022). But while virtual influencers offer unique benefits like content control, they also create potential risks. Only 15% of people describe virtual influencers as trustworthy, in part because they know virtual influencers are not human and thus cannot experience products in real life (Chowdhary 2019; Real Researcher 2022). Our results suggest that what virtual influencers post can help.  Sharing a picture while being with companions, for example, should make them seem like real people, which should increase consumer trust, engagement, and have other downstream benefits for brands that work with them.

Limitations and Future Research

As with any new area, there are a number of interesting questions for future research. First, work might explore other nonverbal cues creators can use to encourage perceptions of anthropomorphism. Humans are uniquely capable of speech, so voice may be one feature that is particularly useful (Schroeder and Epley 2016). That said, with some exceptions (i.e., Lil

Miquela and Lu do Magalu), most virtual influencers do not appear to be using voice, so this may be worth further exploration.

Second, individual differences might also be worth exploring. When people interact with nonhuman agents they perceive as humans, for example, they feel less socially excluded (Mourey, Olson, and Yoon 2017). Consequently, individual differences in need for social belonging might amplify companions' effect by allowing consumers to fulfill social needs. Also, individuals who tend to feel a lack of control over the objects they interact with are more affected by anthropomorphism (e.g., Epley, Waytz, and Cacioppo 2007; Waytz et al. 2010). People tend to be more familiar with how humans (vs. nonhumans) behave, so the simple act of anthropomorphism serves to render these objects seemingly more predictable and therefore controllable (Chen, Sengupta, and Adaval 2018). Future research might investigate whether such people are more affected by companion presence and how it impacts psychological well-being.

Third, beyond companion presence, further research could explore *how* to depict others with a virtual influencer. In the context of selfies, for example, brand selfies (i.e., branded products held by an invisible consumer) result in greater purchase intentions than consumer selfies (i.e., visible consumer faces together with a branded product, Hartmann et al. 2021). Along these lines, in the case of virtual influencers, images could depict them posing with a companion or performing an action together, and the companion could be the same or different gender. Performing an action, for example, might be more beneficial as it suggests the influencer is more real.

Fourth, while we focused on images, the interaction between images and text also deserves attention (Berger et al. 2020). Language cues like humor (Schanke, Burtch, and Ray 2021), or interactive style (Köhler et al. 2011), for example, may also serve a "humanizing" role. Consequently, they may interact with companion presence to shape anthropomorphism perceptions. Similarly, the congruence between text and images may impact the way consumers process information, so future work might want to examine how to effectively blend the two.

Finally, work might also examine other features that have different effects for virtual vs. human influencers. Ancillary analysis of study 1, for example, suggests that positive language might deserve attention. While human influencer posts that use more positive language receive higher engagement, such language seem to backfire for virtual influencers. One reason might be how it is interpreted. Because virtual influencers seem less genuine to begin with, followers might be more likely to interpret positive language (e.g., "this book is *amazing*") as a persuasive attempt, and react against it. Concreteness might be another feature to consider. Results suggest that concrete language has a much more positive effect among virtual influencers. Given concreteness suggests direct experience (Packard and Berger 2021), it might make seem more like the virtual influencer has actually experienced what they are talking about.

Conclusion

In conclusion, this work demonstrates that including others in a picture can have important consequences for consumer attitudes and behavior. By deepening understanding around virtual influencers, we shed light on the effect of visual content in social media, and consumer interaction with non-human agents more broadly.